\documentclass[reprint,amsmath,amssymb,aps,prl,floatfix]{revtex4-1}
\usepackage{graphicx}
\usepackage{dcolumn}
\usepackage{bm}
\usepackage{amssymb}  
\usepackage[none]{hyphenat}
\usepackage[bookmarks=true]{hyperref}
\usepackage{color}

\begin{document}

\title{Evidence for coexisting shapes through lifetime measurements in $^{98}$Zr}
  
\author{Purnima Singh$^1$}
\author{W. Korten$^1$}
\author{T. W. Hagen$^2$}
\author{A. G\"{o}rgen$^2$}
\author{L. Grente$^1$}
\author{M.-D. Salsac$^1$}
\author{F. Farget$^3$}
\author{E. Cl\'{e}ment$^3$}
\author{G. de France$^3$}
\author{T. Braunroth$^4$}
\author{B. Bruyneel$^1$}
\author{I. Celikovic$^{3,5}$}
\author{O. Delaune$^3$}
\author{A. Dewald$^4$}
\author{A. Dijon$^3$}
\author{J.-P. Delaroche$^6$}
\author{M. Girod$^6$}
\author{M. Hackstein$^4$}
\author{B. Jacquot$^3$}
\author{J. Libert$^6$}
\author{J. Litzinger$^4$}
\author{J. Ljungvall$^7$}
\author{C. Louchart$^1$}
\author{A. Gottardo$^8$}
\author{C. Michelagnoli$^8$}
\author{C. M\"{u}ller-Gatermann$^4$}
\author{D. R. Napoli$^8$}
\author{T. Otsuka$^{9,10,11,12}$}
\author{N. Pillet$^6$}
\author{F. Recchia$^{13}$}
\author{W. Rother$^4$}
\author{E. Sahin$^2$}
\author{S. Siem$^2$}
\author{B. Sulignano$^1$}
\author{T. Togashi$^{10}$}
\author{Y. Tsunoda$^{10}$}
\author{Ch. Theisen$^1$}
\author{J. J. Valiente-Dobon$^8$}

\affiliation{$^1$Irfu, CEA, Universit\'{e} Paris-Saclay, F-91191 Gif-sur-Yvette, France}
\affiliation{$^2$Department of Physics, University of Oslo, Oslo, Norway}
\affiliation{$^3$Grand Acc\'{e}l\'{e}rateur National d'Ions Lourds (GANIL), CEA/DRF-CNRS/IN2P3, Boulevard Henri Becquerel, 14076 Caen, France}
\affiliation{$^4$Institut f\"{u}r Kernphysik, Universit\"{a}t zu K\"{o}ln, K\"{o}ln, Germany}
\affiliation{$^5$Institute of Nuclear Sciences “Vinca``, University of Belgrade, Belgrade, Serbia}
\affiliation{$^6$CEA, DAM, DIF, 91297 Arpajon, France}
\affiliation{$^7$CSNSM, IN2P3, Orsay, France}
\affiliation{$^8$INFN, Laboratori Nazionali di Legnaro, Legnaro, Italy}
\affiliation{$^9$Department of Physics, University of Tokyo, Hongo, Bunkyo-ku, Tokyo 113-0033, Japan}
\affiliation{$^{10}$Center for Nuclear Study, University of Tokyo, Hongo, Bunkyo-ku Tokyo 113-0033, Japan}
\affiliation{$^{11}$RIKEN Nishina Center, 2-1 Hirosawa, Wako, Saitama 351-0198, Japan}
\affiliation{$^{12}$Instituut voor Kern- en Stralingsfysica, KU Leuven, B-3001 Leuven, Belgium}
\affiliation{National Superconducting Cyclotron Laboratory, Michigan State University, East Lansing, Michigan 48824, USA}
\affiliation{$^{13}$Dipartimento di Fisica e Astronomia ''Galileo 
Galilei``, Universit\`{a} degli Studi di Padova and INFN Padova, I-35131 
Padova, Italy}

\newcommand\RDDS {Recoil-Distance Doppler-Shift}
\date{\today}

\begin{abstract}
The lifetimes of first excited 2$^+$, 4$^+$ and 6$^+$ states in $^{98}$Zr were measured with the \RDDS~method in an experiment performed at GANIL. Excited states in $^{98}$Zr were populated  using the fission reaction between a 6.2 MeV/u $^{238}$U beam and a $^{9}$Be target. The $\gamma$ rays were detected with the EXOGAM array in  correlation with the fission fragments identified in mass and atomic number in the VAMOS++ spectrometer. Our result shows very small B($E2;2_1^+\rightarrow0_1^+$) value in $^{98}$Zr thereby confirming the very sudden onset of collectivity at $N=60$. The experimental results are compared to large-scale Monte Carlo Shell model and beyond mean field calculations. The present results indicate coexistence of two additional deformed shapes in this nucleus along with the spherical ground state. 
\end{abstract}

\maketitle

The study of various modes of excitations and the associated evolution of nuclear shapes along spin and isospin axes in atomic nuclei is one of the fundamental quests in nuclear physics. While nuclei with ``magic numbers'' of protons and/or neutrons have spherical ground states, as one moves away, the polarizing effect of added nucleons leads to deformation. Throughout the nuclear landscape, this onset of deformation is usually a gradual process, however in neutron rich nuclei around mass $A\sim100$ the shape change is rather drastic and abrupt. 
The ground states of Sr and Zr isotopes with $N$ ranging from the magic number $N = 50$ up to $N<60$ are weakly deformed, however,
they undergo a rapid shape transition from nearly spherical to well deformed prolate deformations as $N = 60$ is approached. 
The sudden nature of  shape transition in Sr and Zr isotopes is evident from the abrupt changes in the two neutron separation energies \cite{Hager06}
and mean-square charge radii \cite{Buchinger,Campbell}, but also from the excitation energies of 2$^+_1$ states and $B(E2)$ values \cite{nndc}.
On the other hand, in 
isotopes with $Z\geq42$ the shape change is rather gradual \cite{Charlwood, Hager06} showing also characteristic signatures of triaxiality. 
This strong dependence of the observed spectroscopic properties, both on the number of protons and neutrons, makes the neutron-rich $A\sim100$ 
nuclei an excellent mass region for testing various theoretical models.

Many 
experimental and theoretical studies have already been reported on the structure of these nuclei.  
More specifically for
the Zr isotopes, the onset of deformation at $N=60$ has been described by a number of theoretical models \cite{GCM,Rodriguez2010,Delaroche,Reinhard99,Skalski,sm2,sm3,sm4,Petrovici2012,Ozen2006,Garcia2005,VanIsacker,Nomuraibm,Xiang2012},
however, none of the models have been able to successfully reproduce the aforementioned rapid change.
Very recently, the abrupt shape changes were correctly described by large-scale Monte-Carlo Shell Model (MCSM) calculations 
\cite{Otsuka, Togashi}.
In the so-called type-II shell evolution scenario, 
the (prolate) deformed states in the isotopes with $N\geq60$ are associated with 
proton excitations to 
the $0g_{9/2}$ orbital.
Driven by the central and tensor components of the effective (proton-neutron) interactions,
these excitations result in a lowering and subsequent filling of the neutron $0g_{7/2}$ and $0h_{11/2}$ orbitals \cite{Togashi}. 
Therefore, both protons and neutrons act coherently to induce the deformation.
These multiparticle-multihole excitation based prolate deformed  states are expected to coexist as excited 0$^+$ state along with the spherical ground state in Zr isotopes with $N<60$.
The crossing of these two distinct coexisting quantum configurations at $N = 60$ manifests as the abrupt change in structure of the ground state and has been interpreted as a quantum phase transition (QPT) from a spherical to a deformed phase \cite{Togashi}. 

Appearance of low-lying $0_2^+$ states in Sr and Zr isotopes with $N<60$ supports the shape-coexistence scenario in these isotopes as an explanation of the rapid shape evolution. However, in order to have a complete picture of the phenomena it is essential to have a precise information on the deformation as inferred from the electromagnetic transition strengths for nuclei in the region of the shape transition.
Experimental programs pursued in the past to determine these parameters in the $A\sim100$ region nuclei have been able to furnish useful insight into the phenomena \cite{Clement,ClementPRC,Chakraborty2013,Kremer}.  For example, coexistence of highly deformed prolate and spherical configurations have already been established in $^{96,98}$Sr \cite{Clement,ClementPRC} and $^{94,96}$Zr \cite{Chakraborty2013,Kremer}. 

Shape-coexistence has also been suggested in $^{98}$Zr \cite{Wu2004,heyde2011}, a key nucleus for understanding the QPT phenomenon in the Zr isotopes. Several attempts have been made in the past to determine the electromagnetic transition rates in $^{98}$Zr. In previous work by Bettermann {\em et al}  \cite{Bettermann}, using the $\beta\gamma\gamma$ fast-timing method, 
a lifetime of $\tau$(4$^+$) = 29 $\pm$9 ps 
was obtained for the 4$^+$ state in $^{98}$Zr, while only upper limits could be obtained for the 2$^+$ state  ($\tau$(2$^+_1$)$\leq$ 15 ps) and the 
6$^+$ state ($\tau$(6$^+_1$) $\leq$ 14 ps).
More recently, in the work by Ansari {\em et al.} \cite{Ansari}, the 
so far most precise
upper limit for the lifetime of $2^+_1$ state in  $^{98}$Zr was 
obtained as $\tau(2^+_1)$ $\leq$ 6 ps, and an upper limit of 15 ps was reported for the 4$^+_1$ state. 
These limits for the 2$^+$ state did not allow to prove whether the onset of collectivity at $N=60$, i.e. in $^{100}$Zr, is as rapid as expected from the drop in excitation energy, 
and predicted by the MCSM, or whether the B(E2) value already increases in $^{98}$Zr. Moreover, in the absence of precise information on $B(E2)$ values one cannot draw any definite conclusion about the properties of the coexisting structures. 

The present letter reports the results of lifetime measurement of the first excited 2$^+$, 4$^+$ and 6$^+$ states in $^{98}$Zr using the \RDDS\ method and confirms for the first time shape-coexistence in this nucleus. In addition, our result shows very small B($E2;2_1^+\rightarrow0_1^+$) value in $^{98}$Zr thereby confirming the very sudden onset of collectivity at $N=60$.

The excited states in $^{98}$Zr were produced in fission reactions, where a $^{238}$U beam with an energy of 6.2 MeV/u delivered by the GANIL facility,
impinged on a  2.3 mg/cm$^{2}$ thick $^{9}$Be target. The VAMOS++ \cite{Pullanhiotan2008,Rejmund2011} spectrometer positioned at 20$^\circ$ with respect to the beam axis was used to detect and identify the reaction products in mass, charge, and atomic number on an even-by-event basis.  
Gamma rays emitted from the reaction products were detected using the EXOGAM array \cite{Simpson2000} consisting of ten segmented Clover detectors arranged in two rings (three detectors at 135$^\circ$ and seven detectors at 90$^\circ$) with respect to the spectrometer axis. 
Correlation between the prompt $\gamma$ rays detected at the target position and the detection of an ion in the focal plane of VAMOS++ triggered the event building, thus allowing the selection of prompt $\gamma$ rays from isotopically identified fission fragments.

\begin{figure}
\centering
 \includegraphics[clip=true,width=.500\textwidth, angle=0]{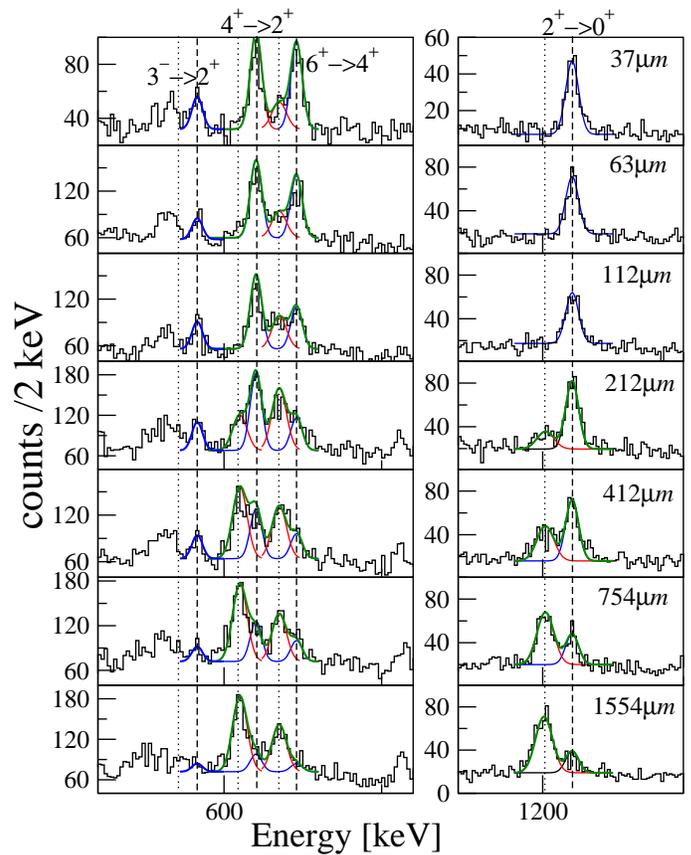}
\caption {\label{fig2} $\gamma$-ray spectra showing the $2^+\rightarrow0^+$, $4^+\rightarrow2^+$, $6^+\rightarrow4^+$ and $3^-\rightarrow2^+$ transitions in $^{98}$Zr, observed in the detectors at 135$^\circ$ for seven target-degrader distances. The spectra are Doppler corrected using the velocity measured in VAMOS++ after the degrader. The dotted and dashed lines indicate the positions of the shifted and unshifted components, respectively. Also displayed are the gaussian fits to the spectra (green), and parameters chosen for fitting the shifted (red), and unshifted (blue) components. 
}
\end{figure}
A compact plunger device was used for the lifetime measurements using the 
\RDDS~(RDDS) technique 
\cite{Dewald2012}. The recoiling $^{98}$Zr nuclei had an average velocity of 37.3 $\mu$m/ps and were slowed down in a 4.9 mg/cm$^2$ thick Mg degrader foil placed behind the target.  Data was collected for seven distances ranging from 37 to 1554 $\mu m$ with average running times of approximately 24 hours per distance. 

The energy of the $\gamma$ rays was Doppler corrected event-by-event using the measured velocity and direction of the fission fragments as measured after the degrader foil in VAMOS++.  
Figure \ref{fig2}~shows the Doppler-corrected $\gamma$-ray spectra of $^{98}$Zr observed with the EXOGAM detectors located at 135$^\circ$  for different target-degrader distances. The $\gamma$ rays that were emitted after passing through the degrader appear at the correct transition energy in the Doppler corrected spectra, whereas those emitted between target and degrader are shifted to lower energies when observed at backward angles. Decay curves $I^u_{ij}(x)$/(${I^u_{ij}(x)+I^s_{ij}}(x)$) were constructed from the intensities of the shifted ($I^s_{ij}(x)$) and unshifted ($I^u_{ij}(x)$) components of the transitions as a function of target-degrader distance. For each distance, the normalisation factors (${I^u_{ij}(x)+I^s_{ij}}(x)$) were found to be consistent with the number of ions identified in VAMOS++. This allowed us to use the relative number of ions identified in VAMOS as normalization factor for the weak and contaminated  583 keV transition depopulating the $3^-\rightarrow2^+$ states.
The lifetimes of individual states were determined from the relative intensities of the two components as a function of target-degrader distance
using the differential decay curve method (DDCM) analysis for singles plunger data \cite{Dewald1989}. All observed feeders have been taken into account as per their experimentally observed intensities. The unobserved feeding contribution was subtracted following the procedure explained in Ref. \cite{Dewald1989}. We have also explored independently the errors in lifetime due to relativistic effects, possible deorientation effects  and solid angle effects, however, error due to these effects were found negligible as compared to the experimental uncertainties and are not taken into account in present analysis. 
Further details on the experiment and the analysis techniques 
can be found in Refs. \cite{Hagen2018,Hagen2017}.

As most important result the lifetime 
of the 2$^+_1$ state was determined 
for the first time
as $\tau=3.8\pm$0.8 ps. 
In addition, we were also able to determine the lifetime for the 4$^+_1$ and  the 6$^+_1$ state as 7.5$\pm$1.5 ps and 2.6$\pm$0.9 ps, respectively. The $B(E2)$ values from the present work are listed in table I. 
Our new lifetime measurement 
yields a $B(E2;2_1^+\rightarrow0_1^+)$ value of 2.9(0.6) W.u, thereby confirming the rapid onset of deformation in Zr isotopes only at $N=60$. 

\begin{table}
\caption{\label{table1}Summary of results for the lifetime measurements in $^{98}$Zr.}

\begin{ruledtabular}
\begin{tabular}{ccccccccc}

$I^\pi_i\rightarrow I^\pi_f$ &$\tau_{exp}$ (ps) &$\tau_{lit.}$ (ps)&$B(E2,\downarrow)_{exp} (W.u.)$ \\
     
\hline
$2^+_1\rightarrow0^+_1$     & 3.8(0.8)&$\leqslant6$\cite{Ansari}  & 2.9(0.6)$^a$(0.2)$^b$ \\
$2^+_1\rightarrow0^+_2$     &         &  & 28.3(6.0)$^a$(2.4)$^b$ \\
$4^+_1\rightarrow2^+_1$     & 7.5(1.5)&$\leqslant15$\cite{Ansari} & 43.3(8.7)$^a$(10.8)$^b$ \\
$4^+_1\rightarrow2^+_2$     &         & & 67.5(13.5)$^a$(16.9)$^b$ \\
$6^+_1\rightarrow4^+_1$     & 2.6(0.9)&   & 103.0(35.7)$^a$ \\
 
\end{tabular}
\end{ruledtabular}
\begin{flushleft}
\footnotesize{The errors in the $B(E2)$ values  from the measured lifetimes$^a$ and from the branching ratios$^b$ taken from \cite{Urban2017} are mentioned separately.\\
}
\end{flushleft}
\end{table}


\begin{figure}
\centering
 \includegraphics[clip=true,width=.280\textwidth, angle=-90]{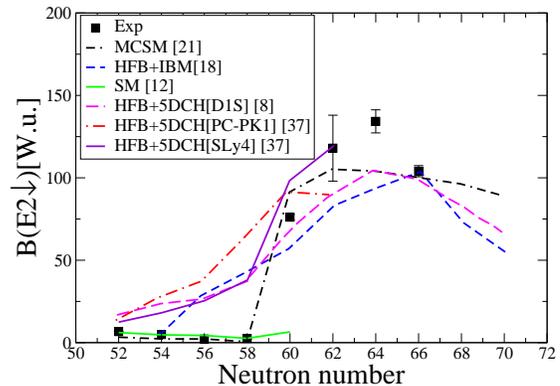}
\caption {\label{fig4}(Color online) Comparison of calculated $B(E2;2_1^+\rightarrow0_1^+)$ values with experimental data for the Zr isotopes. Experimental data points are from refs. \cite{Raman2001,Chakraborty2013,Kumbartzki2003, Ansari,Browne2015plb}. Theoretical values are taken from refs. \cite{Nomuraibm,Mei,sm3,Togashi,Delaroche}.
}
\end{figure}

The onset of deformation in the Zr isotopes has been described by a number of theoretical models. 
Information from $B(E2)$ values on the rapid structural changes in neutron-rich Sr and Zr isotopes is available from calculations
solving a five-dimensional collective Hamiltonian (5DCH) with parameters determined 
within the relativistic mean-field approach
(PC-PK1 force) \cite{Mei}, the non-relativistic Hartree-Fock approach using the SLy4 \cite{Mei} 
and the Gogny-D1S \cite{Delaroche} force. All these calculations suggest a picture of spherical-oblate-prolate shape transition in the neutron-rich Sr and Zr isotopes. 

\begin{figure*}

\centering\includegraphics[clip=true,width=0.310\textwidth,angle=-90]{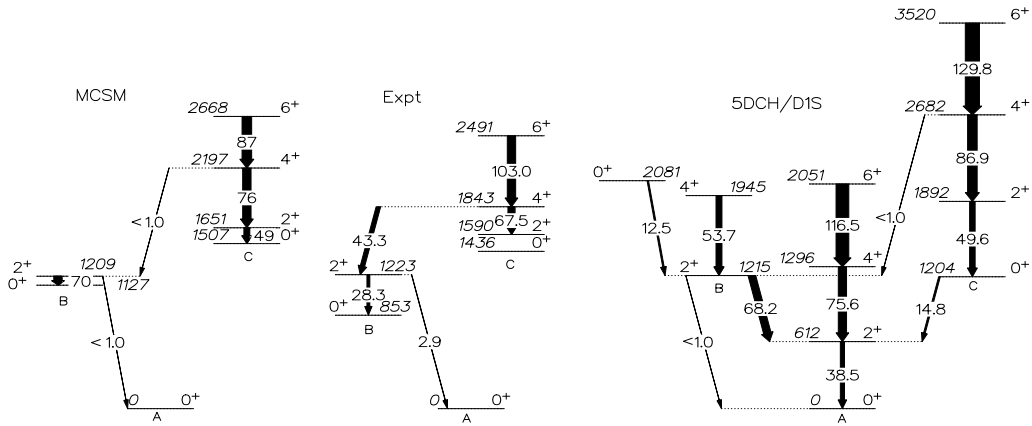} 

\caption{\label{be2_ls} Comparison between the experimental and theoretical level schemes for $^{98}$Zr. The excitation energy and spin are mentioned against each state. Labels and widths of the arrows represent calculated \cite{Delaroche,Togashi} and measured $B(E2)$ values in W.u.}

\end{figure*}

Figure \ref{fig4} compares the $B(E2;2_1^+\rightarrow0_1^+)$ values from these models with the experimental data. It can be clearly noted that the models based on the mean-field approach predict a gradual increase in collectivity already starting at $N\sim54$.
This is also the case for the interacting boson model (IBM) calculations using self-consistent mean-field approximation based on the Gogny-D1M energy density functional \cite{Nomuraibm}. However, certain differences can be noted depending on the underlying effective force. The PC-PK1 and Gogny D1S force predict a very gradual change in most of the collective properties between $N=54$ and $N=60$.
Calculations using SLy4 force, on the other hand,  predict a more pronounced change in collectivity at $N=60$ \cite{Mei}. 
None of them is, however, capable to reproduce the experimental data. 
This could in part be related to the fact that the particle number projection in these methods is only valid on average. 
Simply speaking there might still be components in the theoretical wave functions from the neighboring isotopes with $N\pm2$, which would smoothen out
the evolution with neutron number.
It is also worth noting that the conventional large-scale shell model (SM) calculations \cite{sm3} show a rather accurate reproduction of experimental data up to $N=58$, but are constrained by valence space limitation to $N<60$. 
Finally, the dramatic increase of $B(E2;2_1^+\rightarrow0_1^+)$ values between $N=58$ and 60 is very well reproduced by recent state-of-the-art MCSM calculations
\cite{Togashi}.

Using the most recent branching ratios determined by Urban {\em et al.} in the $\beta$ decay of $^{98}$Y \cite{Urban2017} we were also able to determine the $B(E2)$ values for the decay to the non-yrast states, which give important new information. The experimental excitation energies of the states and the transition strengths between them are compared with the calculations in Fig. \ref{be2_ls}. 
A collective value of B($E2;2_1^+\rightarrow0_2^+$)=28.3(6.4) W.u suggests a common moderately deformed structure for the $0^+_2$ and $2^+_1$ states. Assuming a rigid, axial symmetric deformed shape yields a quadrupole deformation parameter $\beta_2\approx0.21$. These results are in general agreement with the results of MCSM calculations which predict an enhanced B($E2;2_1^+\rightarrow0_2^+$) and very small $B(E2;2_1^+\rightarrow0_1^+)$ transition strengths. However, while the calculations suggest a gradual increase in deformation of the $0^+_2$ states between $^{94}$Zr and $^{98}$Zr \cite{Otsuka}, our present results indicate a deformation which is smaller than that observed in $^{96}$Zr \cite{Kremer}. 

\begin{figure}

\centering\includegraphics[clip=true,width=0.280\textwidth,angle=-90]{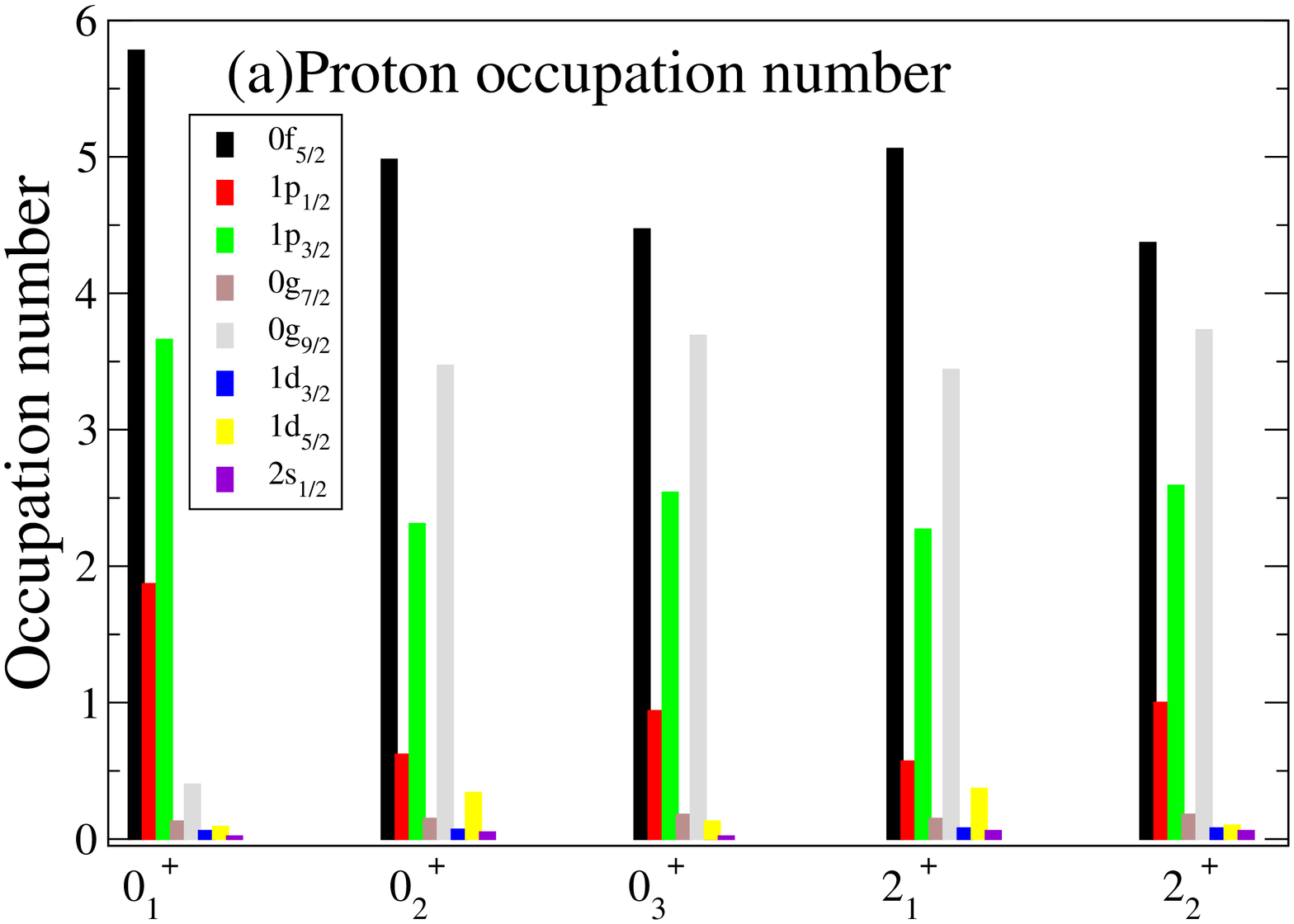}\\
\centering\includegraphics[clip=true,width=0.280\textwidth,angle=-90]{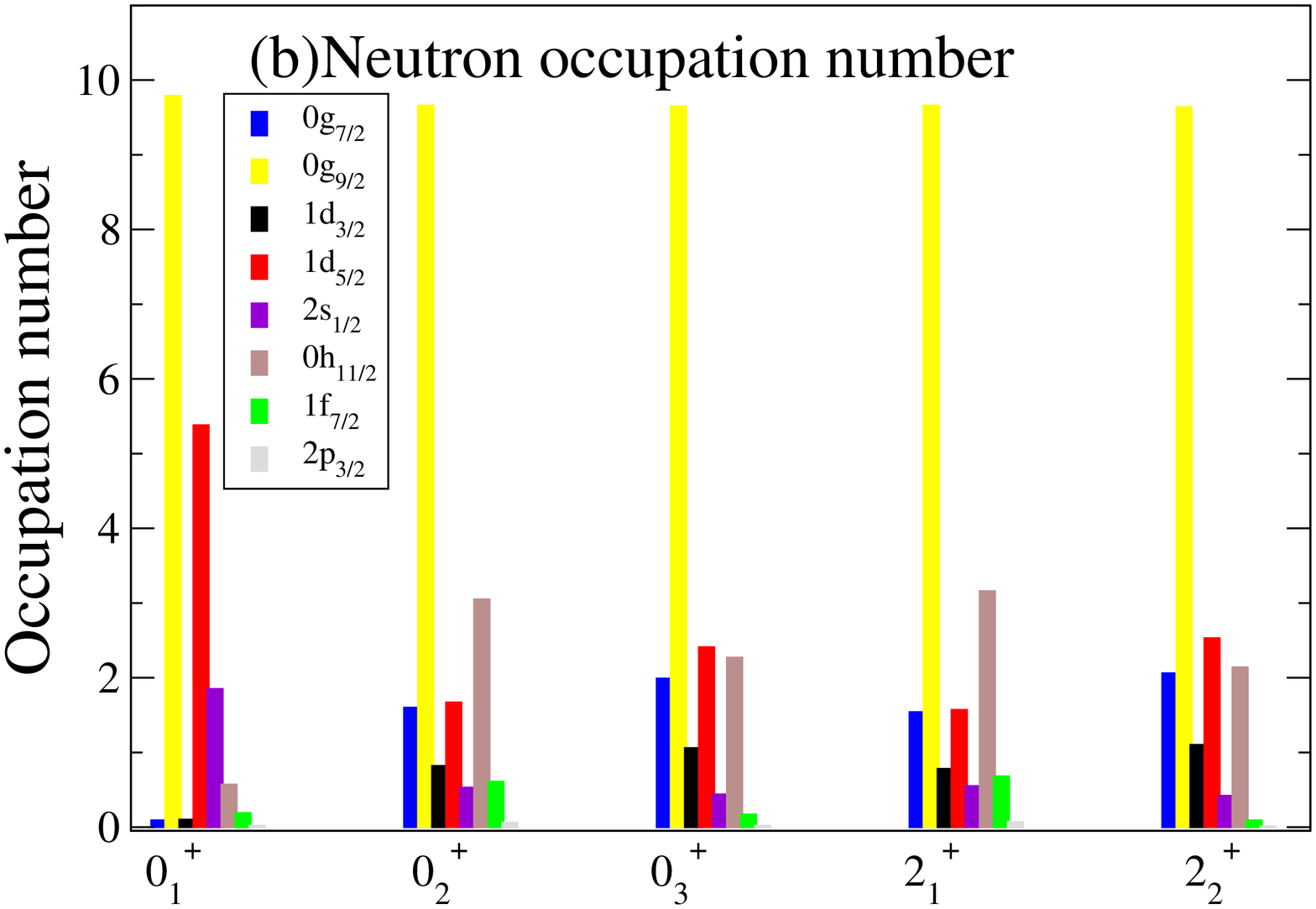} 

\caption{\label{occp}(Color online) Occupation numbers of (a) protons and (b) neutrons for the states in $^{98}$Zr}

\end{figure}

As shown in Fig.~\ref{be2_ls} the $B(E2)$ values in $^{98}$Zr indicate the coexistence of three different structures at low spin:
A nearly spherical 
$0^+_1$ ground state, a moderately deformed excited $0^+_2$ state as discussed above and a well-deformed (band-like) structure  
possibly based on the $0^+_3$ state. 
The existence of latter is supported by the rather large $B(E2)$ values within this band as shown in Fig. \ref{be2_ls}. 
A significant mixing of this well-deformed structure with the moderately deformed configuration, based on the $0^+_2$ state, is indicated by the large  B($E2;4_1^+\rightarrow2_1^+$) value. A strong mixing between the two $2^+$ states was already suggested in ref.~\cite{Wu2004} based on their similar feeding and decay patterns. 
Finally, the large electric monopole transition strength observed between the $0^+_3$ and $0^+_2$ states also support the mixing of two coexisting shapes \cite{heyde2011}.

The proposed triple shape coexistence scenario in $^{98}$Zr is also supported by MCSM calculations, which predict coexistence of three shapes in this nucleus. A spherical $0^+_1$ state, a prolate deformed $0^+_2$ state and a triaxial $0^+_3$ state. The underlying mechanism for stabilizing such coexisting structures has been discussed in terms of type II shell evolution \cite{Togashi}. As shown in Fig. \ref{occp} the calculations, clearly indicate that all the relevant states above $0^+_1$ involve on the average three proton excitations from the $1p0f$ orbitals to the $0g_{9/2}$ orbital and five neutron excitations from the $1d_{5/2}2s_{1/2}$ orbitals to the $0g_{7/2}$, $1d_{3/2}$ and $0h_{11/2}$ orbitals. The structure of the $0^+_2$ and $2^+_1$ states are similar to each other. The $0^+_3$ state and all the other states built on top of it also have similar configurations. The $B(E2)$ values within the band predicted from MCSM calculations are very well in agreement with the experiment for the structure based on $0^+_3$. 
However, it is to be noted that the MCSM calculations 
do not reproduce the strong mixing of the two $2^+$ states (see Fig. \ref{be2_ls}). The observed discrepancies suggest a need for further refinement of the shell model Hamiltonian used in the MCSM calculations.

The five-dimensional beyond mean-field calculations based on the Gogny force (5DCH/D1S) overestimate in general the deformation in $^{98}$Zr. 
These calculations suggest a $K=0$ yrast band structure based on $0^+_1$ state, which gets stretched with increasing angular momentum,
i.e. with mean shape coordinates $\beta$ and $\gamma$ changing from 0.23 to 0.37 and 27$^\circ$ to 16$^\circ$, respectively.
The calculations also predict a second well-deformed prolate band including the levels $0^+_2$, $2^+_3$, $4^+_3$, $6^+_3$, $8^+_3$. 
This band is predicted to be more rigid against triaxiality ($\gamma\sim18^\circ$); furthermore, the mean axial deformation $\beta$ is changing from 0.34 to 0.42.
The transition strengths within the excited prolate band are at low spin rather close to the experimental values of the band built on the 0$_3^+$ state, 
but increase more strongly within the band. 
The deformed structures are therefore reasonably well reproduced by this model, however it fails to reproduce the spherical ground state of $^{98}$Zr, 
a feature which was also observed in other mass regions and also explains the smooth onset of collectivity at $N=60$ observed for all mean field calculations (see Fig.~\ref{fig4}).
%

To summarize, using the \RDDS~method~on isotopically identified fission fragments, we have measured the 
lifetimes in the neutron-rich isotope $^{98}$Zr, which is located just below the predicted quantum phase transition at $N=60$.
The result shows a very small B($E2;2_1^+\rightarrow0_1^+$) value in $^{98}$Zr confirming the very sudden onset of collectivity at $N=60$. 
This effect is well described by the Monte-Carlo Shell-Model and interpreted as Quantum Phase Transition. 
Beyond-Mean-Field calculations are in general not able to reproduce this effect, which could be related to the treatment of the particle number projection. The results from the present measurement confirms coexistence of three distinct shapes in this nucleus for the first time. A comparison of the measured $B(E2)$ values with the state-of-art MCSM calculations indicate spherical-prolate-triaxial shape coexistence in $^{98}$Zr.

The authors thank the technical teams at GANIL for their support during the experiment and the authors of Ref. \cite{Navin} for preparing and optimising the VAMOS++ and EXOGAM spectrometers. We also thank A. Navin for helpful discussions. This work has  received funding from the European Union's Horizon 2020 research and innovation programme under the Marie Sk\l{}odowska-Curie grant agreement No.~702590. It was also supported by the European Community FP7 Integrated Infrastructure Initiative, contract ENSAR No.~262010, by the Research Council of Norway under project grant 213442  and 263030, and by Deutsche Forschungsge-meinschaft (DFG) project DE 1516/3-1. The MCSM calculations were performed on K computer at RIKEN AICS (hp150224, hp160211).
This work was supported in part by the HPCI Strategic Program (The origin of matter and the universe) 
and "Priority Issue on post-K computer" (Elucidation of the Fundamental Laws and Evolution of the Universe) from MEXT and JICFuS.

\end{document}